# Laser-processing of grinded and mechanically abraded Nb-surfaces

V. Porshyn[a)], P. Rothweiler and D. Lützenkirchen-Hecht

*Faculty of Mathematics and Natural Sciences-Physics, University of Wuppertal,*

*42119 Wuppertal, Germany*

## Abstract

The effect of pulsed laser polishing on rough niobium surfaces was investigated. We created different well-defined roughness profiles with standard emery papers and subsequently remelted random surface areas with a size of about 2 x 2 mm$^2$ with ns-laser pulses (wavelength 1064 nm, pulse length 10 ns). Pristine as well as laser-treated surfaces were investigated using optical profilometry and atomic force microscopy, and the surface topography was described by means of correlation functions. Uniformly rough and highly smooth surface geometries were achieved for fractals above and below 7 µm, respectively. Moreover, the behavior of foreign particles during the laser-processing was investigated in detail. The polishing procedure was also monitored point by point by detecting electrical signals, i.e. sample charging, which resulted from the intense laser illumination. The measured electrical charges were found to be correlated with the local surface texture. Thus, regions with initially high roughness profiles and regions with extensive laser-induced defects could be directly identified from the detected electrical signals.

## 1. Introduction

For some years now, laser polishing became an essential part of modern surface processing techniques. There are a number of approaches to smooth surfaces of different materials with intense laser beams. For instance, pulsed or continuous laser polishing as well as a combination of both techniques are widely used [1]. In particular, rough metallic surfaces can be polished to obtain mean

[a)] **Author to whom correspondence should be addressed. Electronic mail:** porshyn@uni-wuppertal.de.

surface roughness values of several hundred nanometers [2-5]. Several studies focus on technically important polishing parameters such as laser fluence, pulse length, spot size, feed rate, overlap, etc. [6-8]. For the successful remelting and polishing of metallic surfaces a set of appropriate parameters can be well estimated by theoretical modelling [9-10]. However, real working parts, which should have a certain surface quality, are mechanically cut out from a piece of raw material and/or first undergo mechanical and chemical treatments. Thus, particulates, residues and roughness features on substantially different length scales are usually present in the pre-treated materials surfaces from the mechanical grinding and polishing, and those impurities may differently respond to a laser illumination. Also, quite different sizes of surfaces defects can significantly affect the optical reflection and absorption of the treated materials, and thereby influence the results of the laser-polishing processes [11]. Thus, a polishing procedure with a fixed set of parameters for such a heterogeneous surface becomes problematic. In contrast, a local adjustment of the parameters point by point is required in order to obtain the desired surface finish. Our previous investigations showed that objects, which are electrically insulated from the ground, can generate a characteristic voltage response, which might be potentially used as an in situ feedback signal during the laser polishing [12]. The reliability of this phenomenon to estimate the post-processed surface quality of a (110) oriented Nb single crystal on a micrometer scale is evaluated in this paper in more detail.

For our study we used niobium, because of its importance for particle accelerators. The high-field performance of superconducting cavities made of niobium strongly depends on impurity concentrations and surface flatness [13]. In particular, local surface defects with jagged geometries like holes, pits, bumps, scratches etc. as well as foreign particles exhibit a reduced field emission threshold and lead to a break-down of superconductivity [14, 15]. The conventional approach to prepare an accelerator cavity involves numerous process steps, e.g. mechanical grinding and polishing, buffered chemical polishing, electro polishing, etc. [16]. These techniques randomly corrode the surface and are not able to quantitatively eliminate all relevant defects [17]. In contrast, surface treatments with lasers can be done point by point locally with high accuracy and without any additional waste or contamination. Thus, they are examined here in more detail in order to evaluate their capabilities and opportunities for the machining of demanding surfaces.

## 2. Materials and methods

### 2.1 Sample Preparation and Characterisation

First, a monocrystalline, (110)-oriented niobium rod with a diameter of 25 mm and a thickness of about 3 mm was mechanically polished to a flat, mirror-like finish. The last polishing step was done with a polishing compound of alumina and wax. The mean size of the used particles was 150 nm in the last polishing step. Typically, the achieved mean surface roughness (Ra) was around 75 nm according to atomic force microscopy (AFM) and optical profilometry (OP) results. Then, the surface was divided into four regions and three of them were roughened with different standard sand papers of a certain, well-defined grit size (see Fig. 2). Finally, the samples were thoroughly cleaned with acetone in an ultrasonic-bath for 10 min. In our study sand papers with designations P500, P1200 and P2000 in accordance with FEPA (Federation of European Producers of Abrasives, https://fepa-abrasives.org/) standard were used, corresponding to a particle size of the emery papers of ca. $30.2 \pm 1.5$ µm, $15.3 \pm 1.0$ µm and $10.3 \pm 0.8$ µm, respectively. The corresponding averaged roughness values as determined by OP for at least five random areas with a size of $1 \times 1$ mm$^2$ per region are presented in Fig. 1a. The roughness increased gradually from 75 nm up to more than 530 nm with a deviation of below 3 % in the major part of the treated regions. The transition zones with a width of about 1 mm between the regions and the outer boundary of the sample were excluded for the surface analysis and the laser treatments here.

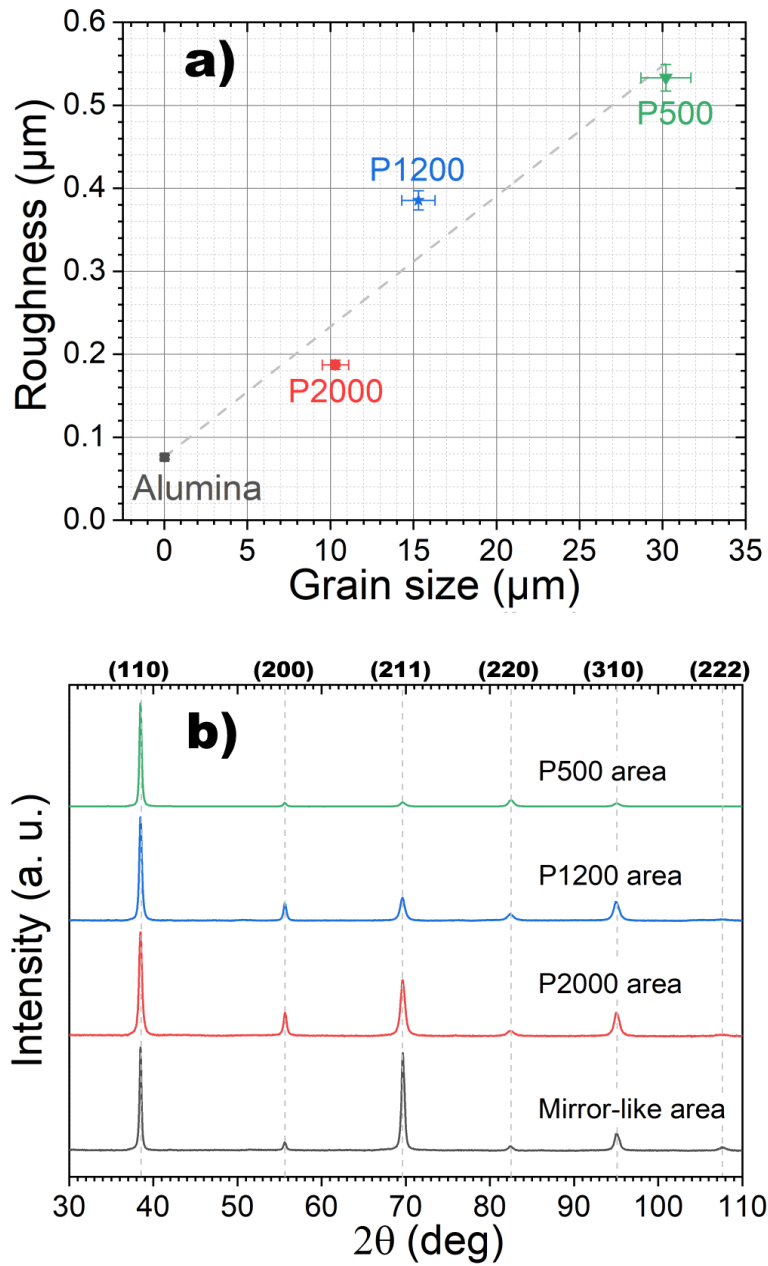

**Fig. 1:** a) Mean average roughness vs. grain size of the used abrasives. The dashed line is meant to guide the eye. b) High resolution X-ray diffraction patterns and the corresponding Miller indices of areas with variable roughness using Cu K-α radiation in the Bragg-Brentano geometry. The data are normalized relative to the dominant reflections from (110) lattice planes.

The X-ray diffraction studies of the abraded surfaces revealed a number of tendencies for the allowed reflections. As shown in Fig. 1b, all possible reflections could be observed due to the defective upper surface layer that is structurally influenced by the mechanical polishing (damage layer). In particular, a relative strong (211)-reflex with a (211)/(110)-intensity ratio of 0.947 systematically appeared in the mechanically prepared mirror-like area. While for a (110) -oriented single crystal, this ratio should be close to zero, the corresponding value for a polycrystalline Nb sample with no random orientation of the crystallites would be about 0.3, indicating a high degree of preferred orientation in the (211) direction for the finest mechanically polished region, see Table 1. The intensity of the second order reflection from (110) increased whereas the intensity of all other peaks decreased with an increasing roughness, i.e. with an increasing X-ray transmission to the monocrystalline, (110)-oriented bulk material through the damage layer [18]. The peak broadening of the dominant (110) reflex increased erratically from around 0.281° for the mirror like area to 0.421° for the P2000 area with a roughness of 187 nm as result of a grain refinement at the interface. A further increase of the roughness subsequently led to a smaller peak broadening of down to 0.319° (P500) following the behavior of the maximum peak intensity, as expected from their mutual relation [19]. In total, taking into account the finite X-ray penetration depth of several micrometers in our case, it can be concluded that the mechanically treated surfaces consist of a certain polycrystalline fraction, which is relevant, in particular, for the optical absorption during the laser polishing process, as well as for the field emission behavior.

|       | **Alumina** | **P2000** | **P1200** | **P500** |
|-------|-------------|-----------|-----------|----------|
| (110) | 16.9        | 17.8      | 25.8      | 67.2     |
| (200) | 8.7         | 23.9      | 26.2      | 13.7     |
| (211) | 47.7        | 28.5      | 17.2      | 7.8      |
| (310) | 16.4        | 24.3      | 27.3      | 11.3     |
| (222) | 10.4        | 5.4       | 3.5       | 0        |

**Table 1:** Fraction of crystallographic orientations of the mechanically prepared niobium surfaces relative to the ideal polycrystalline niobium powder. The accuracy of the data is ±0.1 %.

*2.2 Laser-polishing technique*

The sample was positioned inside a vacuum chamber with a base pressure of 2 x $10^{-4}$ Pa, being electrically isolated from the chamber walls. A pulsed q-switched Nd:YAG laser (Q1B-10-1064 from Topag, Germany) with a wavelength of 1064 nm, a pulse repetition rate of 10 Hz, a pulse duration of 10 ns and a pulse to pulse instability below 1 % was used as an illumination source in our experiments. The beam was focused and step by step scanned over the surface by a motorized plan convex lens with a focal length of 100 mm, see Fig. S1 [38]. The beam spot size and the beam profile were set to be around 250 µm and flat-top, respectively. The polishing procedure and data acquisition were performed with a personal computer and home-made software. The voltage drop on an external resistor induced by the laser pulses was systematically monitored using a fast oscilloscope (PicoScope 4226, Pico Technology Ltd., St. Neots, UK) with 50 MHz-bandwidth and a sampling rate of 125 MS/s. In additional, high temporally resolved measurements of single pulses were carried out with a 2 GHz-bandwidth oscilloscope (Infiniium 54852A, Agilent, Colorado Springs, USA). Finally, surface regions with different sizes in the $mm^2$ range were laser treated as presented in Fig. 2. Optimal polishing results using constant processing parameters were achieved with a pulse number of 80 per single spot, a laser fluence of $(1.55 \pm 0.15)$ $J/cm^2$ and an overlap of 75 % between neighboring points of the treatment. These values were generally applied, unless otherwise noted.

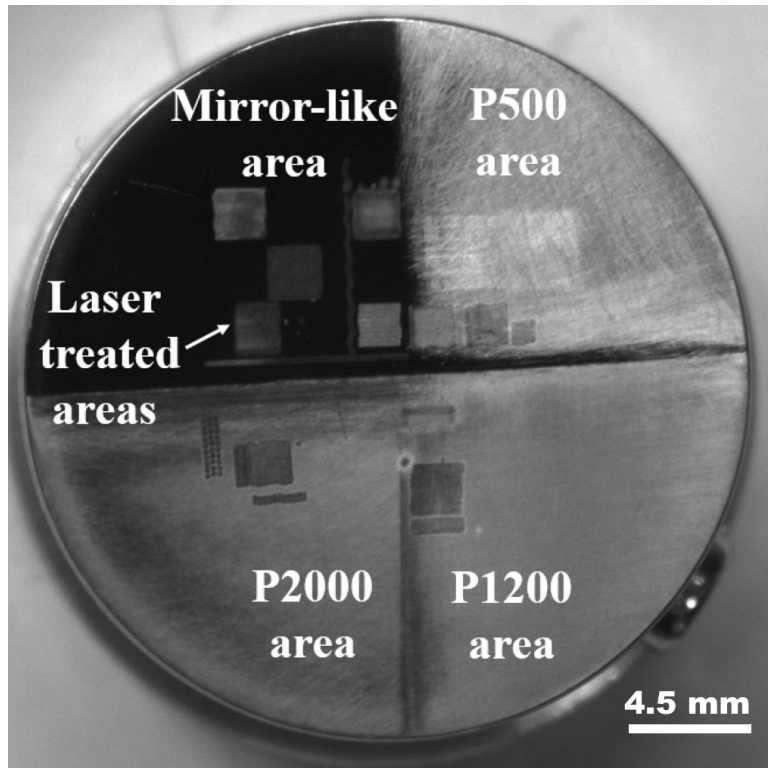

**Fig. 2:** Top view of the niobium sample with a diameter of around 25 mm. The surface exhibits four main regions with different roughness profiles: mirror-like (polished with alumina), P2000, P1200 and P500 as indicated. Rectangle casts with a size of few millimeters are various laser polished areas. Representative areas with a maximum size of 1×1 mm$^2$ outside the transition zones between different pre-treated regions were considered for the numerical analysis.

*2.3 Electric probe study*

In the following, a typical time-resolved voltage response after 20 pulses each of (1.55 ± 0.15) J/cm$^2$ is presented in Fig. 3a. The signal shows three pronounced peaks at 9 ns, 20 ns and 176 ns, respectively. The first peak results from the interaction of the last pulse of the pulse train of the incident laser beam with a heated surface and a metallic vapor, which forms during the intense illumination [20, 21]. After the pulse termination, the next shot increases the electron emission as systematically detected. The nature of this second feature is not quite clear and will be a subject of our future research. The third broad peak at 178 ns can be well related to the formation and the subsequent recrystallization of the melt pool in the surface [22, 23]. During this third period an intense emission of electrons from the liquefied metal occurs [24]. This process can be compared with the thermionic electron emission from a hot solid filament. The phenomenon is highly efficient and can be described by the Richardson–Dushman equation [25]:

$$I = AT^2\exp(-W/kT), \tag{1}$$

with the Boltzmann constant k, the temperature T, the work function W and the universal constant A, which is equal $4.51 \times 10^5$ $A^2/(m^2K^2)$ for niobium. The maximum emission current $I_{max}$ expected from a hot spot on a niobium surface can be estimated using W = 4.37 eV [26], the melting point temperature $T_m$ = 2741 K [27] and an emitting area with a diameter of 250 µm (i.e. the size of the laser spot), resulting in a value of 1.535 mA. Typical emission peaks with a potential drop of around 14 mV were detected over a 50 Ω resistor as presented in Fig. 3, corresponding to an emission current of 0.28 mA, which is substantially lower than estimated using eq. (1).

For our purposes the integral charge per pulse was calculated, the behavior of which was found to be correlated with the surface morphology as presented below in more detail. Moreover, the multi-pulsed operation was essential to achieve a well-defined surface finish. The magnitude of the electron emission from pulse to pulse varied strongly as shown in Fig. 3b in dependence on the pulse number. The most intense excitation of charge was generated by the very first pulses. The next pulses subsequently generated lower charges down to a certain minimum of around (30 ± 8) pC. The lowest charge values were reached after an application of four pulses on the mirror-like surface and after approx. 50 pulses on the rough ones. The presented trend is related to the polishing regime, which results in a smooth finish without laser-induced defects.

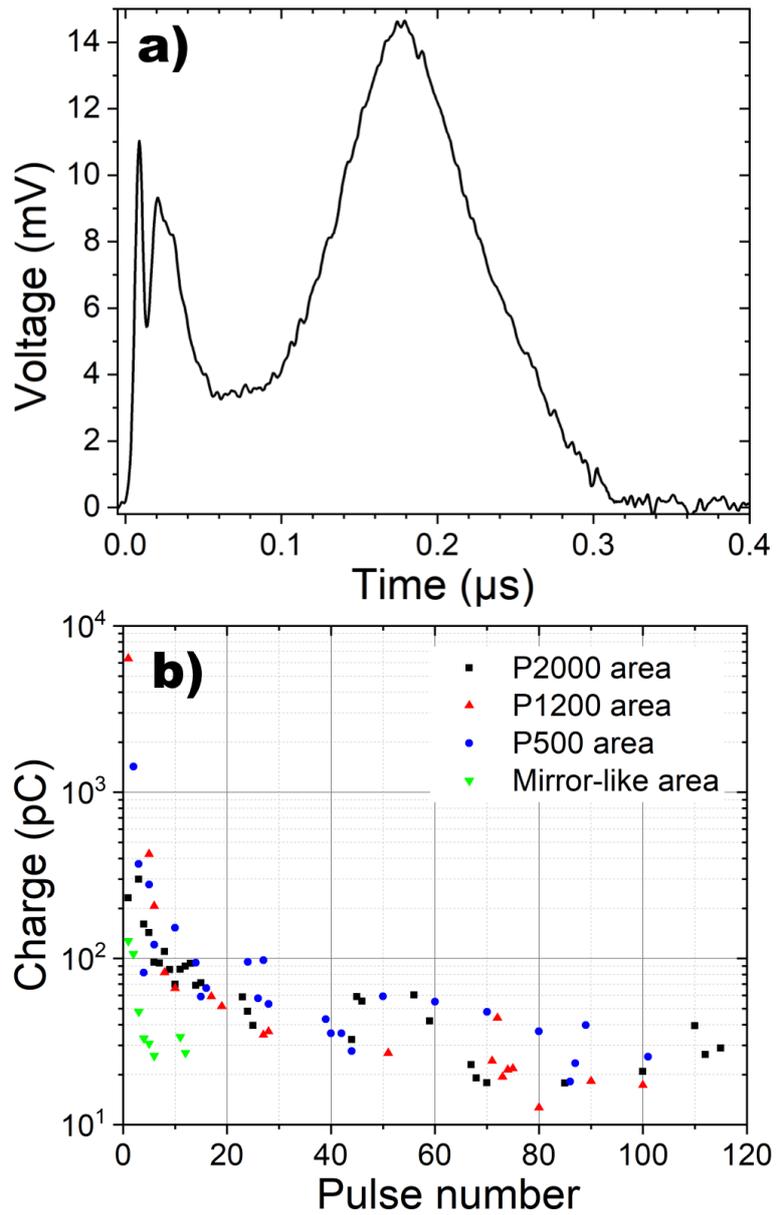

**Fig. 3:** a) Time resolved voltage drop over a 50 Ω resistor as induced by the last single laser pulse of a pulse train of 20 pulses (10 Hz, (1.55 ± 0.15) J/cm$^2$ per pulse). The presented measurement was done on the surface with the mirror-like finish. b) Charge accumulation per pulse for a certain spot in the various regions (mirror-like, P2000, P1200, P500) as subsequently acquired for each pulse in a pulse train.

## 3. Results and discussion

*3.1 Surface morphology*

The effect of our laser polishing on random surface defects is representatively demonstrated in Fig. 4. Here the area initially treated with the sand paper of P500 type sand paper was completely modified. Heavy scratches and edges were smoothed or entirely removed. Instead, complex periodic wave-like structures appeared. This also includes additional laser induced defects in the form of dendritic structures, which will be discussed later. For the analysis of the surface topographies as measured using optical profilometry and atomic force microscopy, a description of the surfaces by means of height-height correlation functions was employed, the latter being fully characterized by the root-mean-square surface roughness and the lateral correlation length [28]. Moreover, the Hurst parameter (H) and the fractal dimension (D) can be extracted from the slopes (β) of log-log graphs of the power spectral density (PSD) and the wave number k (k = 2π/x with x as a point in real space), i.e. $\log(PSD(k))$ vs. $\log(1/k^{\beta})$, to determine the type of surface roughness. In the sense of the fractal hypothesis the following definition applies for the fractal dimension D of a given surface: $2D = 7-\beta$ with $2 \leq D \leq 3$ [29,30].

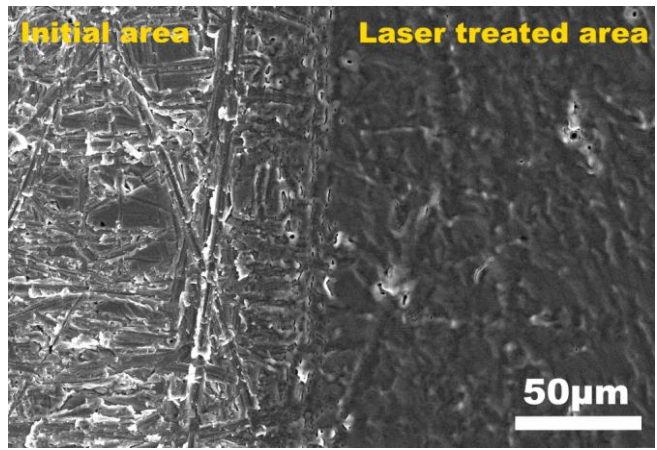

**Fig. 4:** Scanning electron microscopy (SEM) micrograph of a transition region between the initial and the laser processed P500 area.

In the following, the PSD of representative regions were measured with the optical profilometer with a size of 1 x 1 mm$^2$ and additionally using atomic force microscopy with a size of 50 x 50 µm$^2$ for all samples. The obtained results are compiled in Fig. 5 and Fig. 6, respectively. We used a standard calculation method, which includes first the determination of the one-dimensional (1D) PSD for each line of the experimental data along the fast scan axis of the respective instrument and then an averaging over all lines. Due to the finite resolution of the profilometer used for the optical measurements (Fig. 5), a cut off wave number ($k_{max}$) around 2 µm$^{-1}$ resulted, while the resolution of AFM measurements (Fig. 6) at larger k was only limited by white noise.

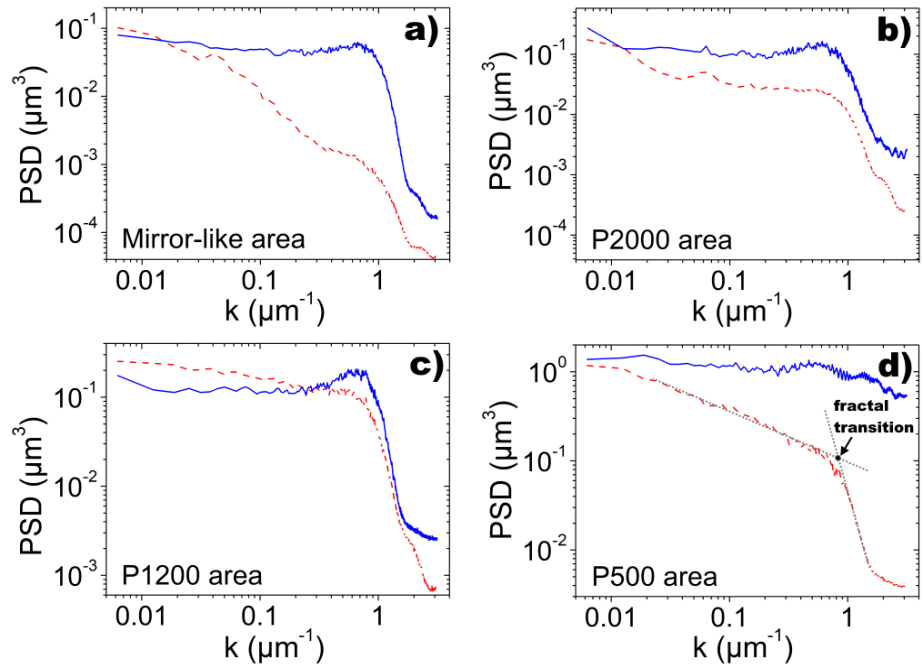

**Fig. 5:** Power spectral densities (PSD) of the initial surfaces (dashed red lines - - -) with mirror-like (a), P2000 (b), P1200 (c) and P500 shapes (d) in comparison to the laser-treated surfaces (solid blue lines ——) as measured with optical profilometry. The transition between two fractals is highlighted by the intersection point $k_{trans}$ between the slopes of the PSD (dotted gray lines ····).

The PSD functions on the millimeter scale, see Fig. 5, introduce a non-linear behavior with two fundamentally different fractals for the initial surfaces in comparison to those after the laser treatment.

The transition wave number $k_{trans}$ between the fractals for the initial surfaces was estimated from the intersection point between two straight lines, see Fig. 5(d), to be 1.48±0.01 µm$^{-1}$ and 0.89±0.01 µm$^{-1}$ for the mirror-like and the abraded areas, respectively. The latter value corresponds to a wavelength of about 8 µm in real space, which might be an evidence for a correlation between the observed fractal transition and the dimensions of the applied abrasive particles of the grinding papers, see Fig. 1a. The slightly different behavior of the mirror-like surface can be explained by its non-uniformity due to additional randomly distributed holes with a depth of approx. 150 nm and lateral dimensions of several microns, see Fig. S2 [38]. Furthermore, the slopes of the log-log graphs for the rough areas in the k-range below 0.78 µm$^{-1}$ increased with an increasing size of the abrasives (see Fig. 5(b)-(d)). The type of created roughness on the mirror-like surface and the abraded surfaces can be well related to the fractional Brownian noise and Gaussian noise with the Hurst parameters: $H_B = (\beta-1)/2$ and $H_G = (\beta+1)/2$, respectively [31]. It should be noted that a Hurst parameter $H \approx 1$ represents a rather smooth and wavy surface, and values close to zero ($H \approx 0$) are characteristic for surfaces with abrupt height changes on small length scales (jagged surfaces). The characteristic values of $H_G$ were found to range between 0.62 for the pristine and 0.76 for the grinded/polished Nb-surfaces as compiled in Table 2.

The laser treated regions exhibit a broad peak in the PSD spectra, which gradually shifts from a k-value of 0.66 µm$^{-1}$ for the mirror-like area to 0.54 µm$^{-1}$ for P500 area. The spectral peak positions correspond to wavelengths between 9.5 µm and 11.6 µm. These values are related to laser-supported absorption waves, which depend on the interaction time of the laser pulse with the liquid metal, as it was shown in our previous studies [12]. The interaction time can be estimated to be between 2–3 ns, in good agreement with the laser pulse duration. Probably, the interaction time slightly increased with an increased roughness. The Hurst parameters ($H_G$) for k < 0.25 µm$^{-1}$ were found to be related to the fractional Gaussian noise as presented in the Table 2. In total the laser polished regions appeared to be more uniform and closer to an ideal white noise value with $H_G = 0.5$ at the micrometer scale in comparison to the untreated ones.

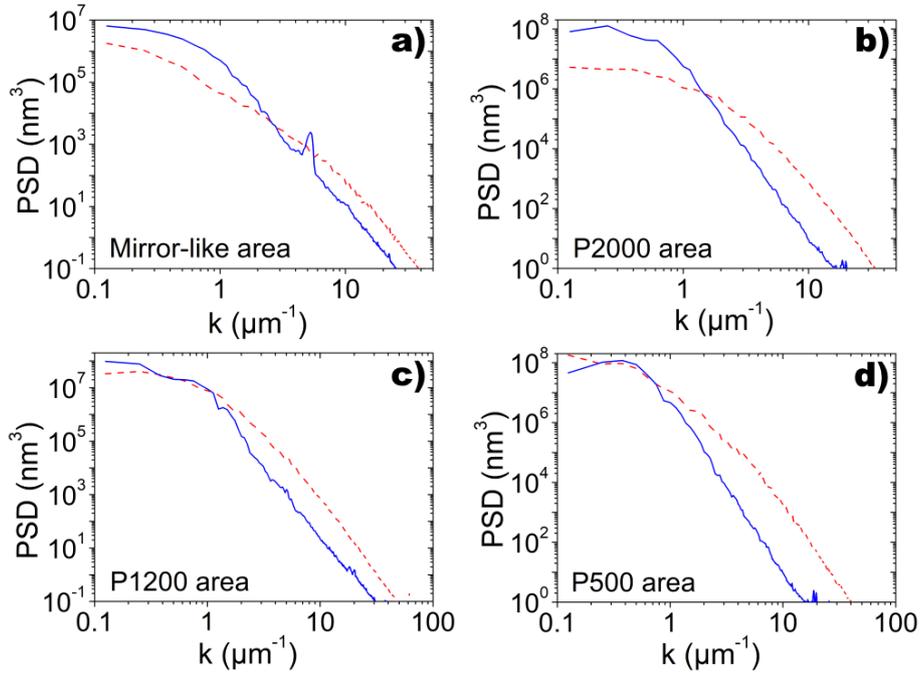

**Fig. 6:** Power spectral densities (PSD) of the initial surfaces (dashed red lines - - -) with mirror-like (a), P2000 (b), P1200 (c) and P500 shapes (d) in comparison to the laser-treated surfaces (solid blue lines ——) as measured with atomic force microscopy.

The surface analysis at the micrometer and sub micrometer scale is presented in Fig. 6. First, it is important to note that the PSD on microscopic scales reveals an entirely different behavior compared to the macroscopic mm scale. The PSD data of the initial surface above the transition wave number exhibits rather non-linear behavior on the log-log scale, i.e. there is no clear linear decay visible for any of the investigated samples. Thus, only a rough estimate of the H-parameters for k above 5 µm$^{-1}$ is possible. This is a fingerprint of a highly non-self-affine surface structure [32], which might consist of non-repetitive frozen individual defects like holes, pits, sharp edges etc. In contrast, the laser treated regions clearly follow a power-law and show scaling factors β between 4.7 and 5 for k well above 1 µm$^{-1}$. These values can be related to highly smooth surfaces with a rigid periodicity, in opposite to the white noise [33]. Thus, we can define the Hurst parameter for the created atomically smooth surfaces to be $H_A = (\beta-4)/2$. Thereby, the laser polished surfaces appear to be smoother compared to the initial ones as presented in Table 2. An additional peak at k ≈ 5.3 µm$^{-1}$ for the mirror-like pre-polished sample directly corresponds to the applied laser wavelength. The amplitude of the waves was found to be around 20 nm, see Fig. S3 [38]. Taking into account the fact that the mean surface roughness of the initial and the laser treated areas become almost equal to approx. 550 nm for the P500 area, see Fig. S4 [38], the effective melting depth induced by the laser processing could be estimated to be well below 1 µm, in good agreement with common theoretical predictions [8, 34].

|  | Alumina | P2000 | P1200 | P500 |
|---|---|---|---|---|
| **Fractal, k < 0.89 µm$^{-1}$** | | | | |
| **Pristine, prior to laser polishing** | | | | |
| D | 2.71±0.01 | 3.00 | 3.00 | 3.00 |
| $H_B$ or $H_G$ | 0.29±0.01 | 0.67±0.03 | 0.62±0.01 | 0.76±0.01 |
| **After laser polishing** | | | | |
| D | 3.00 | 3.00 | 3.00 | 3.00 |
| $H_G$ | 0.60±0.01 | 0.57±0.01 | 0.51±0.01 | 0.54±0.01 |
| **Fractal, k > 0.89 µm$^{-1}$** | | | | |
| **Pristine, prior to laser polishing** | | | | |
| D | 2.00 | 2.00 | 2.00 | 2.00 |
| $H_A$ | 0.38±0.04 | 0.72±0.04 | 0.80±0.05 | 0.66±0.04 |
| **After laser polishing** | | | | |
| D | 2.00 | 2.00 | 2.00 | 2.00 |
| $H_A$ | 0.91±0.02 | 0.88±0.02 | 0.81±0.05 | 0.90±0.03 |

Table 2: Compilation of D- and H-parameters as deduced from the PSD analysis of pristine, polished Nb surfaces and laser treated regions.

*3.2 Energy-dispersive X-ray analysis*

The used polishing abrasives were in particular silicon carbide (SiC) and alumina ($Al_2O_3$) particles. Thus, traces of these materials are generally expected to be present also on the final, mirror-like surface. Traces of alumina were detected neither before nor after the laser treatment. SiC material was detected in holes, which have already been mentioned. After the laser treatment, SiC particles were found to diffuse into subsurface regions of the niobium surface, as presented in Fig. 7. Here, energy dispersive X-ray (EDX) spectra were measured for a smooth surface region (spot no. 1) as well as for defects (spot no. 2 & 3). A comparison of the obtained EDX-data shows that flat regions consist of pure niobium with strong contributions at 2.16 keV and 2.3 keV, and some weaker signals at 0.17 keV, 0.41 keV and 1.91 keV, whereby defective structures like holes or protrusions include intense contributions of silicon at about 1.74 keV photon energy. Pure SiC particles exhibit a high Si/C peak ratio of about 26, see Fig. S2 [38]. The carbon peaks at about 0.28 keV measured for the laser polished sample are substantially larger with respect to the Si signals, and therefore an additional carbon source is evident. We conclude that the detected carbon and oxygen peaks (at about 0.52 keV) are related to residual gases adsorbed on the Nb surfaces. From the measured spectra, we are unable to determine if and to what extent the SiC-particles dissociate in niobium and eventually form e.g. Nb-Si alloys. However, the local character and the shape of the protrusions speak against a possible chemical reaction between the involved elements, and rather support the presence of single particles from the sand papers used. The measured height of the protrusions with Si content was up to 600 nm. No traces of foreign materials other than adsorbed carbon and oxygen were found on the laser treated P2000, P1200 and P500 areas, probably, due to a higher roughness of these regions or a more intense remelting and diffusion of the particles.

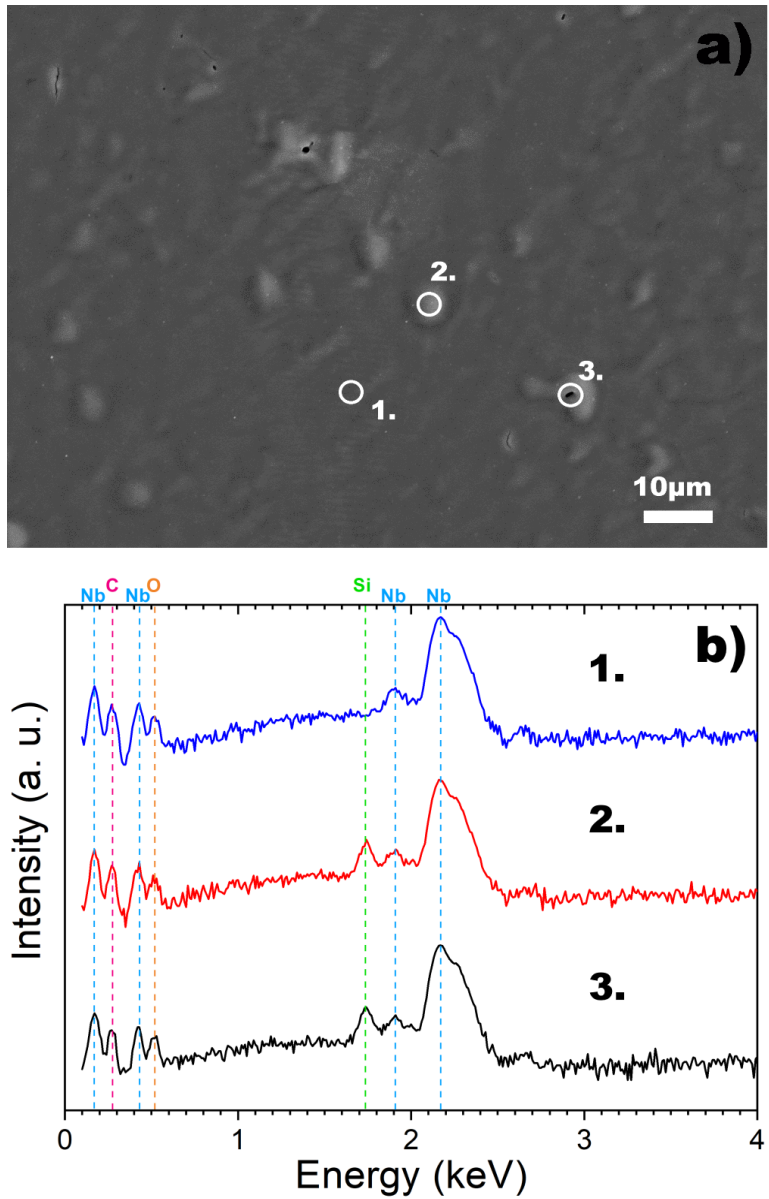

**Fig. 7:** a) SEM micrograph of a laser treated region in the mirror-like area. A highlighted smooth region (spot no. 1) and two different protrusions (spots no. 2 & 3) were analyzed with EDX. b) Energy dispersive X-ray spectra of the corresponding spots in a). The intensity is presented on a logarithmic scale, and the spectra are vertically shifted for a better comparison. The leading peaks are marked in accordance with the associated elements [35].

*3.3 Charge distribution*

The mean charge per single step was measured simultaneously during the laser processing. Independent of the pre-treatment (i.e. mirror-like, P2000, P1200 and P500) it was observed that in all areas the charging of the sample was enhanced by the excessive presence of local defects, which were either originally present on the surface or generated by the laser illumination. The charge values showed the maximum dynamic for the P500 area as presented in Fig. 8a, where a surface with an initial roughness of 533 nm was treated with the laser. According to the acquired charge distribution, the region between 10 and 110 steps in y-scanning direction exhibits more surfaces defects in comparison to the rest of the sample surface. Horizontal lines along the y-scanning direction are an evidence of the scanning pattern. The highest values up to 500 pC were measured in the central part of the investigated area, where striking surface defects appear. The comparison of the charge distribution (Fig. 8a) and the optical image (Fig. 8b) clearly reveals the strong correlation between the surface morphology and the measured electrical response. Moreover, the charge distribution exhibits a kind of mirror symmetry about the Y axis, which shows how sensitively the optics may affect the polishing procedure if fixed parameters are used.

A more detailed view on these defects revealed local cellular structures with a spacing (d) of approx. 10 µm, a height (h) of approx. 30 µm and a column radius of several micrometers, see Fig. 8c and Fig. S5 [38]. The shape of these structures is characteristic for laser induced defects, which can be created on mirror-like niobium surfaces by increasing the laser fluence up to the laser ablations threshold [12]. The created cells appear to absorb the laser light more efficiently compared to other regions, in particular compared to smooth surfaces. It can be directly followed that their melting threshold is significantly lower compared to the applied laser fluence of $(1.55 \pm 0.15)$ J/cm$^2$. This issue resulted in the cellular solidification mode. The transition from the planar to the cellular solidification can be characterized by the magnitude of the thermal gradient (G) and the solidification speed ($v_s$) [36, 37]. Both parameters can be roughly estimated from our measurements as follows. The thermal gradient follows from the cell height with the formula: $G = \Delta T/h \approx 82$ K/µm, where $\Delta T$ = 2450 K is the difference between room temperature (i.e. T=300 K) and the melting point of niobium ($T_M$ = 2750 K). The solidification speed should be larger than 10 m/s for a cell-free solidification taking into account typical pulse lengths and melting depth profiles. Thus, a cooling rate $dT/dt = G\, v_s \geq 0.82\times10^9$ K/s is expected to apply for the planar solidification mode. A reduction of the cooling rate might occur due to an extensive plasma formation, which can be avoided in particular by reducing the laser fluence.

Furthermore, the mean charge values of fine polished regions without cellular structures were measured to be $50 \pm 10$ pC per 100 pulses. Values above 100 pC per 100 pulses already indicated the presence of small local defects with dimensions of several micrometers. However, not all increased charge values represented a remaining surface defect, probably because it was smoothed within the course of 100 pulses. For example, the data points around x=20, y=5 and x=140, y=5 in Fig. 8(a) showed charges of up to 370 pC, but a significantly defective structure was finally observed only at the latter position, while polishing at the first position appeared to be as fine as expected.

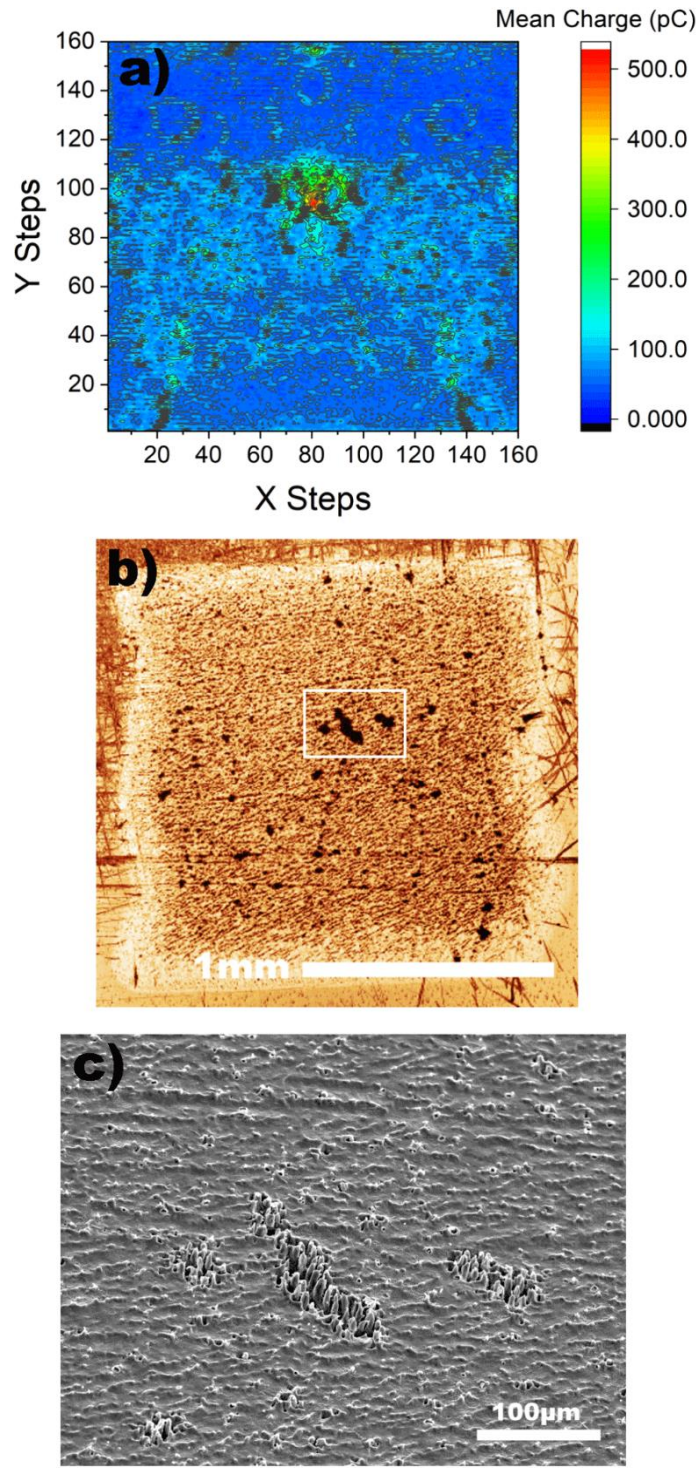

**Fig. 8:** Charge distribution in a) and the corresponding optical image in b) within an area of the laser processed P500 sample acquired during 160×160 steps. c) SEM micrograph of the highlighted area in b), which shows some typical laser induced defects.

# 4. Conclusions

Abraded niobium surfaces were prepared by mechanical grinding and subsequently polished point by point with nanosecond laser pulses. In general, initially random scratches and holes were efficiently removed and replaced by laser induced, smooth, wave-like structures. The characteristic power spectral densities after the laser treatments systematically showed extremely small and high slopes in the log-log scale for the spatial frequencies below and above a certain threshold value of around 0.89 µm$^{-1}$, respectively. This behavior could be related to two different fractal geometries with the extreme fractal dimensions D=2 and D=3. For low frequencies an almost ideal Gaussian noise was created. In contrast, highly smooth fractals with a scaling exponent close to 6 hold for high frequencies. Thus, an additional Hurst parameter was defined to compare the polishing results. Besides of the usual surface geometry a number of additional defects appeared after the laser treatments. It was found that SiC abrasive particles from the grinding papers tend to sink into the bulk and are responsible for the observed protrusions on laser polished mirror-like surfaces. Laser-induced defects also appeared in the form of well-developed cells, which hinted on a reduced cooling rate in some regions with a size of several micrometers. Moreover, the value of the cooling rate, which is necessary to produce a planar geometry, could be estimated to be at least $0.82 \times 10^9$ K/s.

The processing could be monitored by detecting the electrical signals induced by the electron emission from hot spots. The measured signals were well correlated with the surface morphology. A relation between the size of surface defects and the accumulated charge was established, i.e. surfaces with higher roughness emitted more electrons and vice versa. Charge values of around 30 pC for a spot size of 250 µm were found to be optimal for an almost defect free finish of niobium. Regions with extensive laser-induced defects could be directly identified from the charge distributions. However, a precise polishing performance requires an adaptive approach with a variable number of pulses per step. In total, the discussed procedure can be easily integrated into the manufacturing and preparation processes. The application of the described electrical measurements can yield an online, in situ feedback system for the preparation of smooth surfaces, without the need for elaborate, time-consuming post-processing measurements.

**Acknowledgments**

The research was funded by the German Federal Ministry of Education and Research under project number 05H18PXRB1 (STenCiL).